\documentclass[prd,twocolumn,amsmath,amssymb,nofootinbib,floatfix,superscriptaddress]{revtex4-2}

\usepackage{times,graphics}
\usepackage[colorlinks]{hyperref}
\usepackage{makecell}
\usepackage{diagbox}

\usepackage{graphicx}
\usepackage{dcolumn}
\usepackage{bm}
\usepackage{graphicx}
\usepackage{epsfig}
\usepackage{epstopdf}
\usepackage{multirow}
\usepackage{amsmath}
\usepackage{lipsum}
\usepackage{amssymb}
\usepackage{color}
\usepackage{dcolumn}
\usepackage{bm}
\usepackage{natbib}
\usepackage{tabu}
\usepackage{times,graphics}
\usepackage{makecell}
\usepackage{diagbox}
\usepackage{soul,xcolor}

\RequirePackage{mathptmx}      % use Times fonts if available on your TeX system
\RequirePackage{flushend}

\newcommand{\apjs}{{Astrophys.~J.~Supp.}}
\newcommand{\aj}{{Astron.~J.}}
\newcommand{\mnras}{{Mon.~Not.~R.~Astron.~Soc.}}

\def\apj{ApJ}

\def\mnras{MNRAS}

\def\aj{AJ}

\def\apjs{ApJS}

\setstcolor{red}

\graphicspath{{./fig/}}

\begin{document}

\title{Model independent estimation of the cosmography parameters using cosmic chronometers}

\author{Faeze Jalilvand}
\email{f.jalilvand69@gmail.com}
\affiliation{Department of Physics, Bu-Ali Sina University, Hamedan	65178, 016016, Iran\\ f.jalilvand69@gmail.com}

\author{Ahmad Mehrabi }
\email{Mehrabi@ipm.ir}
\affiliation{Department of Physics, Bu-Ali Sina University, Hamedan	65178, 016016, Iran \\Mehrabi@ipm.ir}

\begin{abstract}
Measurement of the universe expansion rate through the cosmic chronometers proves to be a novel approach to understanding the cosmic history. Although it provides a direct determination of the Hubble parameters at different redshifts, it suffers from underlying systematic uncertainties. In this work, we analyze the recent cosmic chronometers data with and without systematic uncertainties and investigate how they  affect the results. We perform our analysis in both model dependent and independent methods to avoid any possible model bias. In the model dependent approach, we consider the $\Lambda$CDM, wCDM and CPL models. On the Other hand,  since the Gaussian process provides a unique tool to study data including a non-diagonal covariance matrix, our model independent analysis is based on the Gaussian process.
\end{abstract}

\maketitle

\section{Introduction}
So far, many efforts have been done to understand the cosmic history. In an isotopic and homogeneous universe, the cosmic history is described by the so called the Hubble function $H(z)$, which is a function of cosmic redshift. In this regard, the observational data is essential to constrain and understand the function.  Except two more futuristic methods namely luminosity dipole \cite{Bonvin:2006en} and redshift drift \cite{Marcori:2018cwn,Lazkoz:2017fvx},  there are mainly two avenues to measure the Hubble function, 1- direct measurements of the function at different redshifts through, cosmic chronometers (CC) \cite{Jimenez:2001gg,Moresco_2012,Farooq:2016zwm,Moresco:2018xdr,Moresco:2020fbm}, or the radial baryon acoustic oscillation (BAO) \cite{Eisenstein:2005su,Percival2010,Blake:2011rj,Reid:2012sw,Abbott:2017wcz,Alam:2016hwk,Gil-Marin:2018cgo}. 2- Constrain through measurement of the cosmic distance.  In the second approach, a standard candle or ruler has been used to measure the comoving distance to a source and the Hubble function can be contained through a relation between the comoving distance and the Hubble function. The most well-known sources are the super novaes (SNs)  \cite{Riess1998,Perlmutter1999,Pan-STARRS1:2017jku}, which give us the luminosity distance at different redshifts. Along with the SNs, the cosmic distance can also be measured through Quasars \cite{Risaliti:2015zla,Risaliti:2018reu}, Gamma-Ray Bursts (GRB) \cite{Demianski:2016dsa,Demianski:2016zxi,Amati:2013sca,Kumar:2014upa}, cosmic microwave background (CMB) \cite{Komatsu2011,Ade:2015yua,Aghanim:2018eyx}, standard sirens \cite{VIRGO:2014yos,LIGOScientific:2020zkf,LIGOScientific:2021aug,CalderonBustillo:2020kcg} and time delay cosmography \cite{Suyu_2010,Unruh:2016adf,refId0,Birrer:2020tax}.

Given a database, there are mainly two viewpoints to constrain the Hubble function. In the first approach, a parametric form has been considered for the Hubble function and the data used to constrain its free parameters. The method depends on the form of the parametrization and different models might give different results specifically at those redshifts which we do not have enough data points. Given a database and a parametric form for the $H(z)$, one can use a statistical tool like the Bayesian inference to constrain the free parameters.  On the other hand, the second approach does not depend on any models or parametrization and the data directly is used to reconstruct the Hubble function. There are various algorithm to perform this task, three most well-known methods are the Gaussian process (GP)\cite{10.5555/1162254},  the Genetic algorithm (GA) \cite{Bogdanos_2009,Nesseris_2010,Vazirnia:2021xuu} and the smoothing method \cite{Shafieloo_2007,Shafieloo_2010,Shafieloo_2018}. Notice that the GP is the most papular approach among all and also is quit flexible in considering data points with a non-diagonal covariance matrix. The method has been used to reconstruct the cosmic history in \cite{Shafieloo_2012,Liao_2019,G_mez_Valent_2018,Pinho_2018,Mehrabi_2020}, tension in the $\Lambda$CDM \cite{Mehrabi:2021cob,Mehrabi:2020zau}, cosmic history from link between background and growth data \cite{Mukherjee:2020vkx,Ruiz-Zapatero:2022zpx} and also the constrain from HII galaxies data \cite{Mehrabi:2021feg}.

The CC approach provides a novel probe to estimate the expansion rate of the universe at different redshifts independent of any cosmological assumptions. The early idea was introduced by \cite{jimenez2002} and rely on the fact that in a isotropic and homogeneous universe the Hubble function is given by $H(z)=-\frac{1}{(1+z)}\frac{dz}{dt}$, where $dt$ is the differential time evolution of the universe at redshift interval $dz$. It is worth noting that according to \cite{Koksbang:2021qqc}, all we need to assume indeed is spatial statistical homogeneity and isotropy.
Since the method does not rely on the functional form of the expansion history or spatial geometry, it provides a unique approach to study the cosmic history. Overall, the CC method depends on the three pillars, 1- definition of a sample as a CC tracers, 2- estimation of the differential age and 3- investigation of the systematic effects. Since the systematic effects are one of the fundamental issues in any cosmological probe like CCs, we consider the latest CC data including all systematics to investigate sensitivity of the model dependent and independent methods to these effects. We consider the CC data with and without systematics, obtain the free parameters of a few well-known models as well as the cosmography parameters to understand how systematics change the final results. Moreover, a similar analysis has been done through a model-independent method to study a possible dependency on the models.

The structure of the paper is as follows, in section (\ref{sec:cc}), we briefly describe the sources of uncertainties in the CC data and provide all essential information regarding CC data analysis. The details of the GP method is given in section (\ref{sec:gp}), also we discuss reliability of the GP in both considering systematic effects and obtaining the derivative of the Hubble function. In addition, we argue about kernel functions in the GP and how it might affect the results. In section (\ref{sec:res}), we discuss the cosmography approach as a reliable method in understanding the cosmic history and then present our results for both model dependent and independent methods. Finally, we provide most important points of our results and conclude in section (\ref{sec:con})

\section{Systematic uncertainties in the cosmic chronometers}\label{sec:cc}
Assuming FLRW universe, the expansion rate of the universe or the Hubble function H(z), is a crucial quantity in understanding the evaluation of the universe. A cosmological probe that can give a direct estimation of the function at different redshift would be very beneficial. As we mentioned above, the radial BAO and CC provide the Hubble function at different redshifts. In contrast to the radial BAO, the CC method is free of any cosmological assumption (except those mentioned in introduction) in determination of data points. However, the systematic effects significantly affect the CC data and a careful investigation is needed to understand the sensitivity of the final results to the systematics. In the following, we briefly mention and describe the systematic effects in the CC data.

In the FLRW framework, the Hubble function is given by
\begin{equation}\label{eq:model}
	H(z) =-\frac{1}{(1+z)}\frac{dz}{dt}.
\end{equation}
To estimate it, one needs to find the differential age of the universe (dt) at a redshift interval (dz). The initial idea was introduced in \cite{jimenez2002}, base on a homogeneous population of astrophysical objects to trace dt, i.e. CC and proposed massive passively evolving galaxies as a sample of optimal CC tracers. There are different methods for obtaining robust estimation of dt from galaxy spectra, including full-spectrum fitting \cite{heavens2000massive,heavens2004star,bothwell2010high,niesche2012emotions} , absorption features (Lick indices) analysis \cite{worthey1994comprehensive,worthey1997hgamma,cardiel2010indexf,borghi2022toward} and calibration of specific spectroscopic features\cite{moresco2012improved}. In general, the CC approach is free of any cosmological assumptions and mainly relies on the age estimates that do not assume any cosmological prior.

The systematic effects might substantially bias the measurement and are divided into four main components. There are some methods to minimize and propagate them to a total covariance matrix.

\textbf{Error in the CC metallicity estimate } An error in the CC metallicity estimate directly affects the measurement of H(z) and consequently its error. In \cite{moresco2020}, this issue has been discussed and authors performed a Monte Carlo simulation of single stellar population (SSP)-generated galaxy spectra considering a variety of stellar population synthesis (SPS) models. They considered the metallicities in the ranges $(\pm 10\%,5\%,1\%)$ and estimating the Hubble parameter. In such a framework, the dependency of the H(z) error due to different stellar metallicity has been investigated.

\textbf{Error in the CC  due to star formation histories (SFH)} The entire SFH which is concentrated in a single burst presents a systematic uncertainty to the CC data. This typically has a contribution of the order of $2–3\%$ and only affects the diagonal terms of the covariance matrix.

\textbf{Assumption of SPS model} This is the major source of the systematic uncertainty in the CC method, no matter which process has been considered to estimate dt. This item introduces non-diagonal elements in the total covariance matrix, as the uncertainties are highly correlated across different spectra. Similar to the above source of uncertainties, the estimation of error on the H(z) due to this item was investigated in \cite{moresco2020}.

\textbf{Rejuvenation effect} If the selected CCs present a residual contamination by a young component, there would be a possible bias due to this issue. According to \cite{moresco2018}, a contamination of $10\%$ ($1\%$) in the total light, by a star forming young component would produces $5\%$ ($0.5\%$) error in $H(z)$ estimation. Notice that this item also add a term to the diagonal part of the covariance matrix.

Other effects, like progenitor bias \cite{franx1996}\cite{van2000} and mass-dependence \cite{moresco2012improved}\cite{moresco20166}\cite{borghi2022toward}, have been shown that have a negligible impact on the estimation of the Hubble function and have not been considered in the current work.

\begin{table}[]
\begin{tabular}{|llll|}
\hline
$z$    & $H(z)$    & $\sigma$ & reference \\ \hline
0.07   & 69.0      & 19.6            & \cite{zhang2014four}          \\
0.09   & 69        & 12              & \cite{simon2005constraints}          \\
0.12   & 68.6      & 26.2            & \cite{zhang2014four}           \\
0.17   & 83        & 8               & \cite{simon2005constraints}          \\
0.179  & 75        & 4               & \cite{moresco2012improved}          \\
0.199  & 75        & 5               & \cite{moresco2012improved}         \\
0.20   & 72.9      & 29.6            & \cite{zhang2014four}           \\
0.27   & 77        & 14              & \cite{simon2005constraints}          \\
0.28   & 88.8      & 36.6            & \cite{zhang2014four}           \\
0.352  & 83        & 14              & \cite{moresco2012improved}         \\
0.38   & 83        & 13.5            & \cite{moresco20166}          \\
0.4    & 95        & 17              & \cite{simon2005constraints}         \\
0.4004 & 77        & 10.2            & \cite{moresco20166}          \\
0.425  & 87.1      & 11.2            & \cite{moresco20166}          \\
0.445  & 92.8      & 12.9            & \cite{moresco20166}          \\
0.47   & 89.0      & 49.6            & \cite{ratsimbazafy2017age}          \\
0.4783 & 80.9      & 9               & \cite{moresco20166}          \\
0.48   & 97        & 62              & \cite{stern2010cosmic}          \\
0.593  & 104       & 13              & \cite{moresco2012improved}          \\
0.68   & 92        & 8               & \cite{moresco2012improved}         \\
0.75   & 98.8      & 33.6            & \cite{borghi2022toward}          \\
0.781  & 105       & 12              & \cite{moresco2012improved}          \\
0.875  & 125       & 17              & \cite{moresco2012improved}          \\
0.88   & 90        & 40              & \cite{stern2010cosmic}           \\
0.9    & 117       & 23              & \cite{simon2005constraints}          \\
1.037  & 154       & 20              & \cite{moresco2012improved}          \\
1.3    & 168       & 17              & \cite{simon2005constraints}          \\
1.363  & 160       & 33.6            & \cite{moresco2015raising}          \\
1.43   & 177       & 18              & \cite{simon2005constraints}          \\
1.53   & 140       & 14              & \cite{simon2005constraints}          \\
1.75   & 202       & 40              & \cite{simon2005constraints}         \\
1.965  & 186.5     & 50.4            & \cite{moresco2015raising}        \\ \hline
\end{tabular}
\caption{$H(z)$ measurements obtained with the CC method and their uncertainties.}\label{tab:hub}
\end{table}

In order to make it clear what data has been used in current work, we have provided all CC data and their uncertainties in Tab.(\ref{tab:hub}). Notice that in this case the covariance matrix is a diagonal matrix so the data is without considering the systematics.  

On the other hand, the systematic effects can be encoded in a non-diagonal covariance matrix. We provide a simple Jupyter notebook in \href{https://github.com/Ahmadmehrabi/Cosmic_chronometer_data}{(CC-systematic)}, which can be used to generate the data with systemic effects. ( for more details on its formalism see \cite{moresco2020}).

It is worth noting that constrain from the CC data has been compared with other cosmological probes like SN Ia and BAO in several works \cite{Moresco:2016mzx,Vagnozzi:2020dfn,Gonzalez:2021ojp,Lin:2019htv}. Contrary, in this work, we consider only the CC data with and without systematic effects and study sensitivity of the model parameters constrain as well as the cosmography parameters on these effects.

\section{Model independent description of data and the Gaussian process}\label{sec:gp}
Given a database and a model including some free parameters, there are some well-known approaches to constrain the free parameters. In fact, the data is used along with a statistical tool like the Bayesian inference to find a constrain on the free parameters. In this scenario, the model prediction might be different from one model to another and the final results depends on the form of the model. On the other hand, a model-independent or a non-parametric approach, tries to infer information using only the data and free of any parametrization or model. In this case, any possible model bias on the final result can be removed. One of  the well-known non-parametric method is the GP which is widely used in the cosmological context \cite{Shafieloo_2012,Liao_2019,G_mez_Valent_2018,Pinho_2018,Mehrabi_2020}.

Now , we briefly introduce the main steps of the GP. Given a database, including $C_D$, the covariance matrix of the data,
\begin{equation}\label{eq:d}
D=\{(x_{i},y_{i},C_D)|i=1,..,n\},
\end{equation}
Our aim is describing the data by reconstructing a function $y= f(x)$ in a model-independent way. The GP is a sequence of Gaussian random variables which can be described by a mean function $\mu(x)$ and a covariance matrix $\Sigma$, so the reconstructed function is given by
\begin{equation}\label{eq:fx}
 f(x)\sim GP(\mu(x),k(x,\tilde{x})),
\end{equation}
where $\tilde{x}$ is an arbitrary point in a domain of the reconstruction and $k(x,\tilde{x})$ is a kernel function for describing the covariance matrix. The most well-known kernel function is the Gaussian kernel,
\begin{equation}\label{eq:gp}
	k(x,\tilde{x})=\sigma^{2}_{f}\exp(-\frac{(x-\tilde{x})^{2}}{2l^{2}}),
\end{equation}
where $\sigma_f$ and $l$ are two hyper parameters which should be constrained by the data. Given the data set and the kernel function, the conditional distribution is then used to predict the function at any new points $x^{\star}$. In this case, the mean and covariance matrix are given by,
\begin{eqnarray}\label{eq:GP}
	\mu^{\star} &=& K(x,x^{\star})[K(x,x^{\star})+C_D]^{-1}Y\\
	\Sigma^{\star} &=& K(x^{\star},x^{\star}) - K(x^{\star},x)[K(x,x^{\star})+C_D]^{-1}K(x,x^{\star}),
\end{eqnarray}
where $Y$ is the column vector of observation $y_i$. The final step is finding the best value of the hyper parameters which can be done by building a likelihood function and then find the best hyper parameters by maximizing the likelihood or Bayesian inference \cite{10.5555/1162254}.

We should emphasize two important points, 1- the scenario allows us to consider a non-diagonal covariance as well as a diagonal one for the covariance of the dataset. This is crucial for consideration of the systematic uncertainties in CC data. 2- the form of the kernel function might affects the final results, so we consider two Matern $(\nu=3.5,\nu=4.5)$ kernels along with the Gaussian to avoid any possible bias due to the GP kernel. To see more details of these kernels see \cite{Mehrabi:2020zau}. It is worth noting that the family of Matern kernels is a generalization of kernel (\ref{eq:gp}) and it is widely used in multivariate statistical analysis. In this case, the absolute exponential kernel is parameterized by an additional parameter $\nu$. If $\nu$ goes to infinity, the kernel becomes Eq.~(\ref{eq:gp}) and if $\nu=1/2$ the kernel becomes equivalent to the absolute exponential kernel.

Another useful property of the GP is that the derivative of the reconstructed function can be easily obtained. In fact, the first and second derivative are given by \cite{10.5555/1162254}:
\begin{equation}\label{eq:fxx}
f^{\prime}(x)\sim GP(\mu^{\prime}(x),\frac{\partial^{2}k(x,\tilde{x})}{\partial x\partial\tilde{x}}),
\end{equation}
and
\begin{equation}\label{eq:fxxx}
f^{\prime\prime}(x)\sim GP(\mu^{\prime\prime}(x),\frac{\partial^{4}k(x,\tilde{x})}{\partial^{2}x\partial^{2}\tilde{x}}).
\end{equation}

Since we are going to compute the cosmography quantities, we need the derivative of the reconstructed function which can be easily  obtained using above formula.

\section{Cosmography in model dependent and independent approaches}\label{sec:res}
As we mentioned above, in a FLRW universe, the Hubble function is a key quantity. The function depends on the content of the universe and is parametrized as,
 \begin{eqnarray}\label{eq:h_par}
 	H(z) &=& H_0[\Omega_m(1+z)^3 + \Omega_r(1+z)^4 + \Omega_k(1+z)^2\\ \nonumber
  &+& 	(1-\Omega_m-\Omega_r-\Omega_k)\Omega_x(z)]^{1/2} ,
 \end{eqnarray}
where $\Omega_m$, $\Omega_r$, $\Omega_k$  are the matter, radiation, curvature density at present time respectively and  $\Omega_x$ is a general function to describe the evolution of the dark energy density. In the current work, we consider a flat universe so $\Omega_k=0$. In addition, since we are investigating the low redshifts data, the radiation term has not been considered. Three well-known models, namely the $\Lambda$CDM, wCDM and CPL have been considered in this work. The dark energy evolution ($\Omega_x(z)$) are given in these models as:
\begin{eqnarray}\label{eq:cpl}
	\Omega_x(z) &=& 1,\\
	\Omega_x(z) &=& (1+z)^{3(1+w_0)},\\
	\Omega_x(z) &=&(1+z)^{3(1+w_0+w_1)}e^{-3w_1\frac{z}{1+z}} ,
\end{eqnarray}
respectively.

We perform a Bayesian inference using Pymc3 package \cite{pymc} to constrain free parameters of these models. The both kind of the CC data (with and without systematic uncertainties) have been used in the analysis. The results are summarized in Tab.(\ref{tab:mcmc}).
In the $\Lambda$CDM model, the data including systematic effects gives $8.2\%$ uncertainty on the $H_0$ while not considering these effects gives $4.7\%$ error.  In this case, the systematic effects increase the $H_0 $ error around $3.5\%$. On the other hand, we obtain $18\%$ and $21\%$ errors on the $\Omega_m$ using two different covariance matrix of the CC data. However, in both cases, all the parameter values are consistent at $1\sigma$ confidence level and we do not see any deviation due to systematic effects.

In the wCDM model, the errors on the $H_0$ are $10\%$ and $12\%$ for two kind of CC data. Moreover, we find $5\%$ ($2\%$) more uncertainty in the $\Omega_m$ ($w_0$) parameter with CC data including systematic effects. Finally, in the CPL model, we also see a similar pattern for the uncertainties of the parameters. Overall, our MCMC results indicate that the systematic effects on the CC data increase the error of the parameters around $2-5\%$.

\begin{table*}[h]
	\centering
	\begin{tabular}{|c|c|c|c|c|c|c|}
		\cline{1-7}
		\multicolumn{1}{|c|}{Model} & \multicolumn{2}{c|}{$\Lambda$CDM}    & \multicolumn{2}{c|}{wCDM}    & \multicolumn{2}{c|}{CPL}     \\ \cline{1-7}
		\multicolumn{1}{|c|}{}                       & \multicolumn{1}{c|}{diag} & non-diag & \multicolumn{1}{c|}{diag} & non-diag & \multicolumn{1}{c|}{diag} & non-diag \\ \hline
		$H_{0}$                                            & \multicolumn{1}{l|}{$67.64\pm3.20$}  & $66.64\pm5.49$  & \multicolumn{1}{l|}{$72.04\pm7.31$}  & $70.25\pm8.16$  & \multicolumn{1}{l|}{$72.88\pm7.47$}  & $71.12\pm8.24$  \\ \hline
		$\Omega_{m}$                                            & \multicolumn{1}{l|}{$0.33\pm0.06$}  & $0.34\pm0.07$  & \multicolumn{1}{l|}{$0.30\pm0.06$}  & $0.31\pm0.08$ & \multicolumn{1}{l|}{$0.32\pm0.08$} & $0.33\pm0.08$ \\ \hline
		$w_{0}$                                           & \multicolumn{1}{c|}{---} & --- & \multicolumn{1}{l|}{$-1.40\pm0.60$} & $-1.45\pm0.64$ & \multicolumn{1}{l|}{$-1.38\pm0.68$} & $-1.48\pm0.72$ \\ \hline
		$w_{1}$                                           & \multicolumn{1}{c|}{---} &  ---& \multicolumn{1}{c|}{----} &  ---& \multicolumn{1}{l|}{$-0.98\pm2.65$} & $-1.13\pm2.69$ \\ \hline
	\end{tabular}
\caption{The best value of the free parameters and their $1\sigma$ uncertainties. }\label{tab:mcmc}
\end{table*}

In addition to the Hubble function, one can define other key quantities base on the derivatives of the Hubble function. These quantities are called the cosmography parameters and the declaration q(z) and jerk j(z) are two important parameters, defined as:
\begin{eqnarray}\label{eq:q-j-h1}
	q(z) &=& (1+z)\frac{H'(z)}{H(z)} - 1,\\
	j(z)& = &(1+z)^2[\frac{H''(z)}{H(z)}+(\frac{H'(z)}{H(z)})^2] - 2(1+z)\frac{H'(z)}{H(z)}  + 1 \label{eq:q-j-h2}.
\end{eqnarray}

In order to investigate how systematic uncertainties affects these parameters, we compute them in different models using both kinds of CC data. In fact, the best value of parameters have been used to find the mean of each cosmography parameters and their uncertainties estimated by the sample in the MCMC chain. The results are presented in Tab.(\ref{tab:cos}).
\begin{table*}[h]
	
	\centering
	\begin{tabular}{|c|c|c|c|c|c|c|}
		\cline{1-7}
		\multicolumn{1}{|c|}{Model} & \multicolumn{2}{c|}{$\Lambda$CDM}    & \multicolumn{2}{c|}{wCDM}    & \multicolumn{2}{c|}{CPL}     \\ \cline{1-7}
		\multicolumn{1}{|c|}{}                       & \multicolumn{1}{c|}{diag} & non-diag & \multicolumn{1}{c|}{diag} & non-diag & \multicolumn{1}{c|}{diag} & non-diag \\ \hline
		
		$H_{0} $    & \multicolumn{1}{l|}{$67.64\pm3.20$}  & $66.64\pm5.49$  & \multicolumn{1}{l|}{$72.04\pm7.31$}  & $70.25\pm8.16$  & \multicolumn{1}{l|}{$72.88\pm7.47$}  & $71.12\pm8.24$  \\ \hline
		
		$q_{0}$      & \multicolumn{1}{l|}{$-0.50\pm0.09$}  & $-0.48\pm0.10$  & \multicolumn{1}{l|}{$-0.95\pm0.66$}  & $-0.97\pm0.69$ & \multicolumn{1}{l|}{$-0.97\pm0.81$} & $-0.95\pm0.82$ \\ \hline
		
		$j_{0} $      & \multicolumn{1}{c|}{$1.0$} & $1.0$ & \multicolumn{1}{l|}{$3.96\pm4.2$} & $4.2\pm4.5$ & \multicolumn{1}{c|}{$3.50\pm6.66$} & $3.15\pm6.83$ \\ \hline
	\end{tabular}
	\caption{The cosmography parameters at present time and their $1\sigma$ uncertainties in different models.}\label{tab:cos}
\end{table*}
Since the $H_0$ parameter has been discussed in the MCMC part, we only discuss the deceleration and jerk parameters in this part.

In the $\Lambda$CDM, the errors on the $q_0$ are $18\%$ and $20\%$, considering diagonal and non-diagonal covariance matrix respectively.  The $q_0$ in wCDM and CPL is close to -1 and has error larger than $\Lambda$CDM. However, all the values are consistent with each other. Regarding the $j_0$, we see a larger value compare to the $\Lambda$CDM in both wCDM and CPL but all values are consistent with the $\Lambda$CDM at $1\sigma$ confidence level since the $j_0$ errors are relatively large.

\begin{table*}[h]
	
	\centering
	\begin{tabular}{|c|c|c|c|c|c|c|}
		\cline{1-7}
		\multicolumn{1}{|c|}{Kernel} & \multicolumn{2}{c|}{Gaussian}    & \multicolumn{2}{c|}{Matern72}    & \multicolumn{2}{c|}{Matern92}     \\ \cline{1-7}
		\multicolumn{1}{|c|}{}                       & \multicolumn{1}{c|}{diag} & non-diag & \multicolumn{1}{c|}{diag} & non-diag & \multicolumn{1}{c|}{diag} & non-diag \\ \hline
		
		$H_{0} $    & \multicolumn{1}{l|}{$67.57\pm4.75$}  & $67.22\pm6.15$  & \multicolumn{1}{l|}{$68.80\pm5.17$}  & $68.25\pm6.49$  & \multicolumn{1}{l|}{$68.60\pm5.05$}  & $68.10\pm6.38$  \\ \hline
		
		$q_{0}$      & \multicolumn{1}{l|}{$0.01\pm0.57$}  & $-0.01\pm0.60$  & \multicolumn{1}{l|}{$-0.14\pm0.69$}  & $-0.16\pm0.72$ & \multicolumn{1}{l|}{$-0.14\pm0.65$} & $-0.15\pm0.69$ \\ \hline
		
		$j_{0} $      & \multicolumn{1}{c|}{$3.4\pm2.13$} & $3.53\pm2.25$ & \multicolumn{1}{l|}{$3.24\pm3.39$} & $3.31\pm3.53$ & \multicolumn{1}{c|}{$3.35\pm2.99$} & $3.56\pm3.12$ \\ \hline
	\end{tabular}
	\caption{The GP results for the cosmography parameters at present time and their $1\sigma$ uncertainties.}\label{tab:gp}
\end{table*}

\begin{figure}[h]
	\centering
	\includegraphics[width=0.5\textwidth]{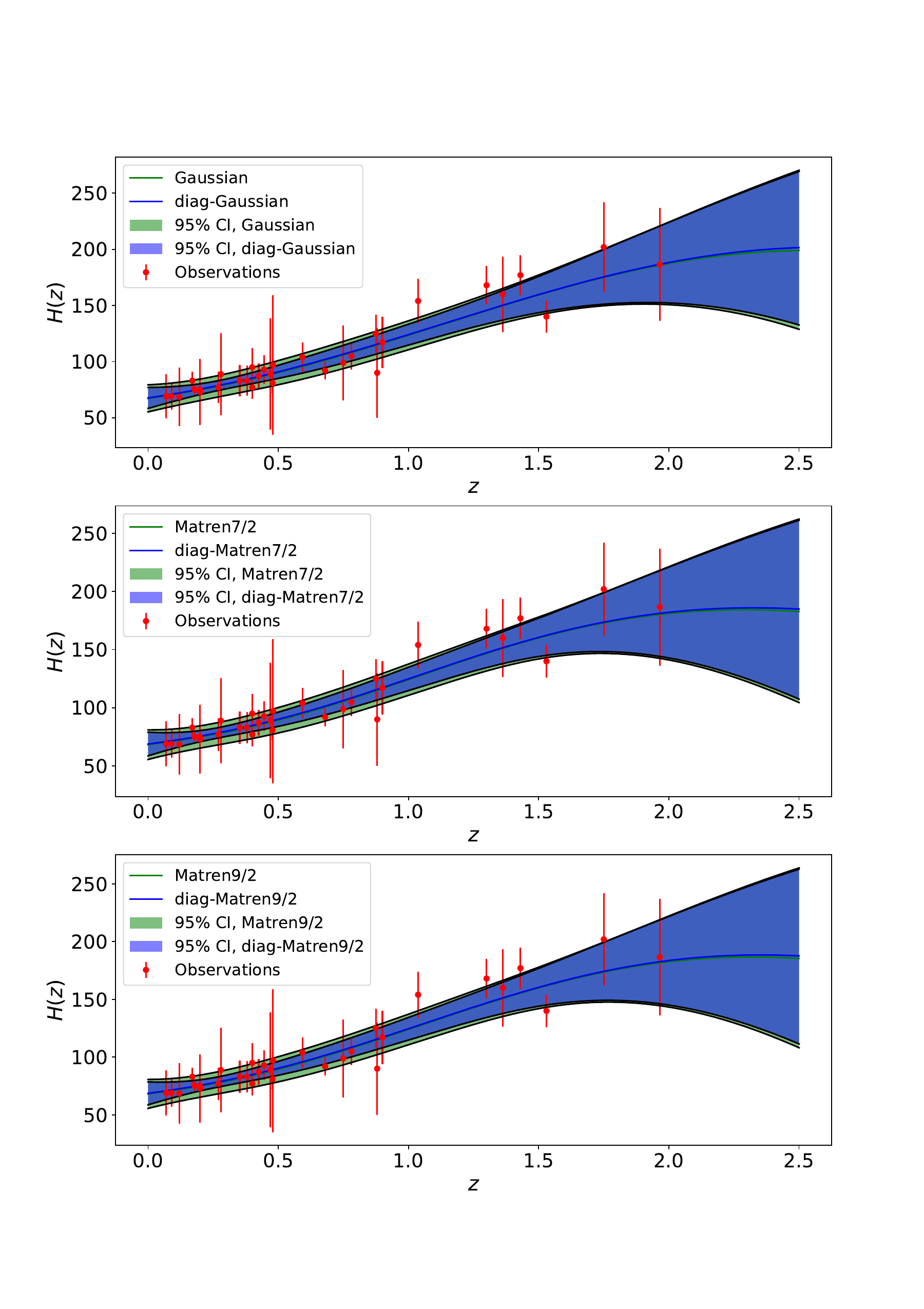}	
	\caption{The GP results of the H(z) using different kernels. The green (blue) shows the results with (without) considering the systematic effects. The color regions indicate the $95\%$ confidence interval. }
	\label{fig:h}
\end{figure}

\begin{figure}[h]
	\centering
	\includegraphics[width=0.5\textwidth]{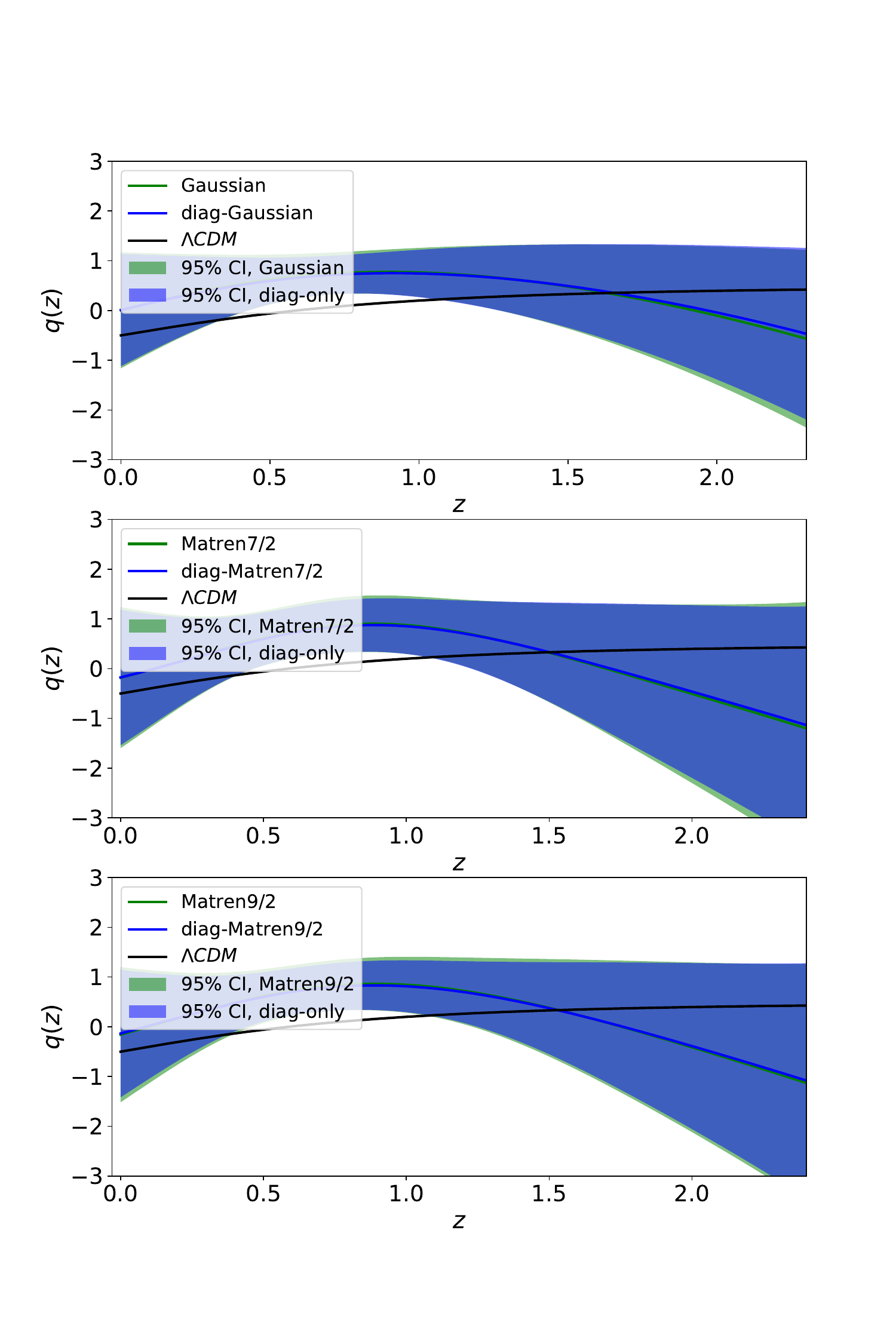}	
	\caption{The GP results of the q(z) using different kernels. The green (blue) shows the results with (without) considering the systematic effects. The color regions indicate the $95\%$ confidence interval.}
	\label{fig:q}
\end{figure}

\begin{figure}[h]
	\centering
	\includegraphics[width=0.5\textwidth]{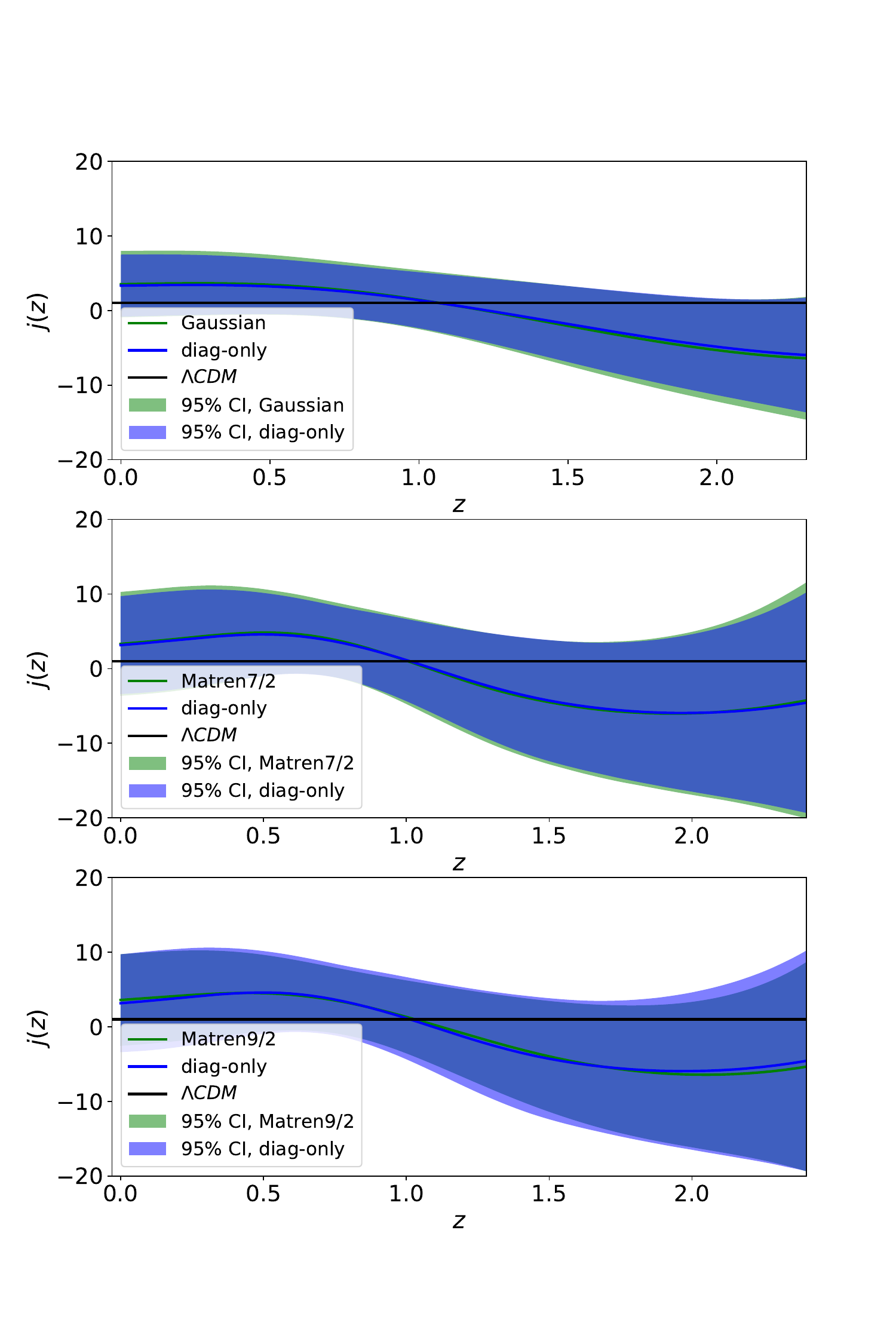}	
	\caption{The GP results of the j(z) using different kernels. The green (blue) shows the results with (without) considering the systematic effects. The color regions indicate the $95\%$ confidence interval. }
	\label{fig:j}
\end{figure}

Moreover, the cosmography parameters have been computed in the GP approach independent of any parametrized model. The results at present time have been shown in Tab.(\ref{tab:gp}). Since the GP kernel might affect the final results, we consider three different kernels namely, the Gaussian, Matern($\nu=7/2$) and Matern ($\nu=9/2$). The $H_0$ results not only are quit good in agreement in different kernels but also in agreement with results of above three models at $1\sigma$ confidence level. In addition, we see a similar pattern in the $H_0$ uncertainties with and without considering the systematic effects. In this case, the $H_0$ error is around $7\%$ ($9\%$) without (with) considering the systematic effects. In contrast to the model dependent approach, the GP gives a $q_0$ close to zero in different kernels but due to a larger uncertainty, all results are consistent with the $\Lambda$CDM. Notice that the $q_0$ errors in the GP are close to those of the wCDM but the CPL model provides a larger uncertainty compere to the GP. Moreover, similar to the model dependent case, we see a tiny difference in uncertainties of $q_0$ with and without considering the systematic effects.

Furthermore, the GP gives the jerk parameter around $3.5$ in different kernels. The values are in agreement with $\Lambda$CDM at $1\sigma$ confidence level. Moreover, we do not see a significant difference with and without considering the systematic effects. This results indicate that the systematic effects mainly affect the $H_0$ and do not change the $q_0$ and $j_0$ significantly.

Finally,  the results of the GP for cosmography parameters at different redshifts have been shown in Figs.(\ref{fig:h},\ref{fig:q} and \ref{fig:j}). The green (blue) line shows the mean value of the parameter with (without) considering systematics and the green (blue) region presents the $95\%$ ($1.96\sigma$) confidence intervals. In each plot the top, middle and bottom panel present the results of the Gaussian, Matern ($\nu=7/2$) and Matern ($\nu=9/2$) respectively. The results of different kernels are consistent with each other and we see no significant difference due to the kernels form. On the other hand, the difference due to the systematic effects is larger at redshifts $z\in(0.2-1)$ compare to other redshifts and mainly affects the Hubble function. For example, at $z=0.5$, the $H(z)$ error is $3\%$ ($6.3\%$) using diagonal (non-diagonal) covariance matrix. In the $q(z)$ and $j(z)$ plots, the black solid line shows the best $\Lambda$CDM model and as it is clear the results are consistent with the model at different redshifts but the $q(z)$ shows a small deviation at redshifts $z\in(0.5-1)$.

\section{Conclusion}\label{sec:con}
Measurement of the expansion rate of the universe at different redshift is crucial to understand the evaluation of the universe.
While CCs provide a promising probe to do this task, their data points are highly affected by the systematics uncertainties. In this work, we use a recent database of the CCs with and without systematic uncertainties to investigate how systematics affect the information gain from CCs. With this aim, the cosmography parameters in both model dependent and independent approaches have been computed using the CC data. In the model dependent case, the $\Lambda$CDM, wCDM and CPL models have been considered, then the best value of free parameters in each model along with their uncertainties have been computed through the Bayesian inference. In the $\Lambda$CDM model, we see a $3.5\%$ ($3\%$) difference on $H_0$ ($\Omega_{m}$) due to the systematic effects. On the other hand, while the uncertainties of the free parameters in the wCDM and CPL are slightly larger than $\Lambda$CDM, the difference due to systematic effects is between $2-5\%$.

Moreover, we use the best value of parameters to obtain the mean values of the cosmography parameters. Our results indicate that the systematic effects highly influence the $H_0$ parameter and difference in $q_0$ and $j_0$ is smaller. In addition, we obtain a smaller $q_0$ (around $-1$) and a larger $j_0$ (around $3-4$) compare to the $\Lambda$CDM results. However, these values are in agreement with the $\Lambda$CDM at 1$\sigma$ confidence level thanks to their large uncertainties.

Furthermore, we use the GP approach to compute the cosmography parameters directly from data and without considering any model. Since the results might depend on the GP kernel, two Matern kernels have been used along with the Gaussian kernel. Overall, the values of all cosmography parameters in all kernels are consistent at $1\sigma$ level with each other. In this case, similar to the model dependent method, the systematics mainly affects the $H_0$ parameter and both $q_0$ and $j_0$ do not change significantly. Moreover, we obtain a smaller value for $q_0$ with uncertainty $\sim 0.6$ which is consistent with results of model dependent method. The GP gives $j_0\sim 3.5$ which is close to the results of wCDM and CPL and show a deviation compare to the $\Lambda$CDM. So our results indicate that the data provides a larger $j_0$, while the $\Lambda$CDM gives $j_0=1$. This is a clear example of the model bias which can be avoided by considering a model independent approach.

Finally, the cosmography parameters as a function of redshift have been computed using the GP method. The results indicate that the systematic effects mainly affect the $H(z)$ at redshift range $0.2-1$ and other parameters do not change significantly. In fact, the difference of $H(z)$ at $z=0.5$ due to the systematics is $\sim3.3\%$ which is very close to the difference in $H_0$. Meanwhile, the $q(z)$ and $j(z)$ are consistent (at $95\%$ confidence level) with the $\Lambda$CDM at all redshifts, except a small deviation in $q(z)$ at redshift range $0.5-1$.

\section{Data Availability Statement}\label{sec:data}

The datasets analyzed during the current study are available in this \href{https://github.com/Ahmadmehrabi/Cosmic_chronometer_data}{(CC-systematic)} Repo on github.
%=================================================================

\bibliographystyle{apsrev4-1}
\bibliography{ref}

\end{document}